\documentclass[conference]{IEEEtran}
\IEEEoverridecommandlockouts
\usepackage[dvips]{graphicx}
\graphicspath{ {images/} }
\usepackage[utf8]{inputenc}
\usepackage[T1]{fontenc}
\usepackage[english]{babel}
\usepackage[shortcuts]{extdash}
\usepackage[noadjust]{cite}
\usepackage{amsmath}
\usepackage{amssymb}
\usepackage{mathtools}
\usepackage{dsfont}
\usepackage{algorithm}
\usepackage{algpseudocode}
\allowdisplaybreaks
\usepackage{tikz}
\usepackage{booktabs}
\usepackage{multirow}
\usepackage{float}
\usepackage[position=top,font=footnotesize,caption=false]{subfig}
\usepackage{xcolor}
\usepackage{url}
\usepackage{verbatim}
\usepackage{float}

\makeatletter

\let\old@ps@headings\ps@headings
\let\old@ps@IEEEtitlepagestyle\ps@IEEEtitlepagestyle
\def\confheader#1{%
    \def\ps@IEEEtitlepagestyle{%
        \old@ps@IEEEtitlepagestyle%
        \def\@oddhead{\strut\hfill#1\hfill\strut}%
        \def\@evenhead{\strut\hfill#1\hfill\strut}%
    }%
    \ps@headings%
}
\makeatother

\confheader{%
        \parbox{\textwidth}{\centering This article has been accepted for publication in the  IEEE International Conference on Sensing, Communication and Networking (SECON Workshops), 2018.}
}

\begin{document}

\title{Robust Multi-Path Communications\\ for UAVs in the Urban IoT}
\author
{\IEEEauthorblockN{\large Zoheb Shaikh, Sabur Baidya and Marco Levorato}
\IEEEauthorblockA{Computer Science Department, UC Irvine, CA, US\\
Email: \{zsshaikh,sbaidya,levorato\}@uci.edu}
\vspace{-8mm}
}

\maketitle

\IEEEpubid{\begin{minipage}{\textwidth}\ \\\\\\[12pt]\\\\ \centering
  \\ \copyright 2018 IEEE. Personal use of this material is permitted. Permission from IEEE must be obtained for all other uses, in any current or future media,\\including reprinting/republishing this material for advertising or promotional purposes, creating new collective works, for resale\\or redistribution to servers or lists, or reuse of any copyrighted component of this work in other works.
\end{minipage}}

\begin{abstract}
Unmanned Aerial Vehicle (UAV) systems are being increasingly used in a broad range of scenarios and applications. However, their deployment in urban areas poses important technical challenges. One of the most prominent concerns is the robustness of communications between the ground stations and the UAVs in a highly dynamic and crowded spectrum. Indeed, competing data streams may create local or temporary congestion impairing the ground stations to control the UAVs.
The main contribution of this paper is a robust multi-path communication framework for UAV systems. The framework continuously probes the performance of multiple wireless multi-hop paths from the ground stations to each UAV, and dynamically selects the path providing the best performance to support timely control. Numerical results, based on a real-world implementation and extensive field experimentation, demonstrate the ability of the proposed framework to provide robust control against exogenous interference and network congestion.
\end{abstract}

\vspace{1mm}
\begin {IEEEkeywords}
Unmanned Aerial Vehicles, Urban Internet of Things, Congestion Control, Network Path Selection
\end{IEEEkeywords}

\section{Introduction}
\label{intro}
Unmanned Aerial Vehicles (UAV) are being increasingly used in a broad spectrum of scenarios and applications~\cite{rodrigues2017uav}.
Their integration in the Urban Internet of Things (IoT) is attracting a considerable interest, for instance to enhance the ability of the city-wide system to support delicate tasks such as surveillance and monitoring, virtual reality, disaster management, and to improve or maintain network coverage.

UAVs are typically controlled by a Ground Control Station (GCS), which wirelessly interconnects with the UAV to build a data-control loop composed of an upstream flow of control messages and a downstream flow of telemetry and sensor data. The urban environment poses several challenges undermining the ability of the GCS to control the UAVs.
First, the topological characteristics of the urban environment may severely limit the operating range due to Line of Sight (LoS) obstruction. This issue has been partially addressed in prior work by creating mesh networks of cooperating UAVs.
However, another important issue that remains largely unaddressed is the coexistence of UAV-related traffic with competing IoT data streams. Exogenous traffic sharing the same access and/or backbone network may create localized and temporary congestion impairing the ability of the GCS to establish an effective data-control loop with the UAV.

This paper addresses these important problems by proposing a robust multi-hop multi-path framework for the remote control of UAV systems. The data and control links are established using the network infrastructure available in urban environments. In particular, we use for communications the 2.4GHz ISM band, which is shared with other Wi-Fi devices and used by other wireless technologies. The multiple paths from the GCS and UAV are continuously probed to quickly select the best option. Importantly, simple local measurements, such as channel sensing and signal strength, would not protect the GCS-UAV communications against local network congestion.

The framework employs a multi-hop multi-path beacon forwarding technique to continuously monitor the performance of the paths from the GCS to the UAV. The UAV measures beacon delay and loss to migrate control routing from one path to another when the current path falls outside of a predefined Quality of Service (QoS) region.

The main contributions of this paper are:
\emph{(a)} A cooperative networking model which establishes multi-hop routes using the urban IoT communication infrastructure to forward control messages from the GCS to the remote UAVs;
\emph{(b)} A framework to dynamically adapt the route used to forward control messages from the GCS to the UAVs based on the current QoS of the paths; and \emph{(c)} A real-world implementation and extensive field experimentation of the proposed framework.
Experimental results show a considerable improvement in terms of control messages reliability, which leads to a reduced delay in accomplishing mission objectives.

The rest of the paper is organized as follows. In Section II, we discuss related work and emphasize the main innovations introduced by this paper with respect to existing frameworks. Section III presents the architecture and describes the adaptive communication and control strategy used to dynamically select the best path from the GCS to the UAV. In Section IV, we describe the experimental setup and provide numerical results assessing the performance of the framework. Section V and VI conclude the paper.

\begin{figure*}[!t]
   \centering
   \includegraphics[scale=.6]{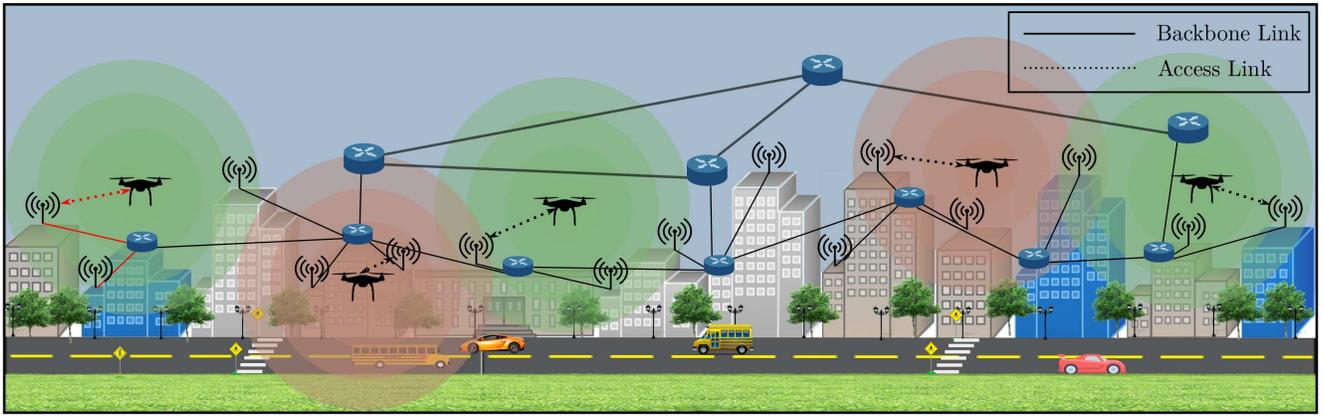}
    \vspace{-0.2cm}
    \caption{Unmanned Aerial Vehicles (UAVs) operating in an urban environment. The dynamics of traffic created by competing applications and the high mobility of the UAVs make robust control challenging. In this paper, we present a multi-hop multi-path adaptive networking strategy to solve those issues.}
    \vspace{-0.4cm}
    \label{fig:uaviot}
\end{figure*}

\section{Related Work}
\label{relatedwork}
Due to the exponentially increasing diffusion of UAVs, the development of effective communication frameworks supporting their operations has received considerable attention in recent years~\cite{bekmezci2013flying}. The interested reader can find in~\cite{gupta2016survey} a detailed survey on the challenges of UAV communications in terms of mobility, fast topology changes, and connectivity. An investigation of IEEE 802.11a applied to UAV-to-ground links can be found in~\cite{yanmaz2011channel}. However, an organic and comprehensive solution to these issues is still missing. 

Related to the methodology used in this work,~\cite{de2010uav} presents a study on UAV systems supporting the connectivity of wireless sensor networks. In~\cite{ha2010uav}, the authors propose an analytical framework to partition the geographical region and maintain a connected graph of UAV nodes. A framework to make UAV networks self-organizing is presented in~\cite{orfanus2016self}. The methodology is based on beacons, whose failure trigger navigation directive to maintain connectivity. Other contributions address the problem of dynamic routing over wireless networks composed of fast moving UAVs, referred to as Flying Ad-hoc Network (FLANET). The solution in~\cite{rosati2016dynamic} extends an existing routing protocol to address ad-hoc networking scenarios.

Most existing approaches assume a dedicated spectrum for UAVs communications.
This paper proposes a framework integrating UAV systems in the Urban IoT using available communication resources to route control messages. A dynamic path selection mechanism ensures robustness against congestion generated by other data streams using the same infrastructure and spectrum. Different from most contributions in this area, we provide a full implementation and experimental investigation.

\section{System Architecture}
\label{sysmodel}

\subsection{Preliminaries}
Fig.~\ref{fig:uaviot} illustrates the scenario considered in this paper: a network of UAVs immersed in an urban environment where a multitude of other sensing and communication devices operate and coexist. 
Due to the topology of urban environments, a direct link between a UAV and GCS would likely fail to provide a satisfactory communication range, with an inevitable drop in the reliability of control messages delivery. Importantly, the UAV incorporates fail-safe mechanisms that are activated when the UAV is disconnected from the GCS, including GPS-based return-to-home function and emergency landing. However, in both cases the UAV fails to accomplish the assigned mission. 

Hence, we use the wireless Access Points (AP) available in the city to forward control messages from the GCS to the UAV and telemetry data back from the UAV. The APs are interconnected through the backbone network with established minimum cost paths calculated using either Link State or Distance Vector protocols. In this paper, we focus on communications in the $2.4$~GHz band using Wi-Fi technology. However, the same reasoning can be applied to any, or multiple, technologies depending on the communication capabilities of the UAVs.

\subsection{Architecture}

Current approaches addressing connectivity in urban environments primarily use Received Signal Strength Indicator (RSSI) to perform AP selection. However, each individual AP and the router involved in the path to the AP may be also supporting other data streams, which may create localized congestion and affect the performance of a subset of the possible paths. Intuitively, messages from the GCS have stringent delay requirements, where excessive delay may affect controllability, or trigger fail-safe mechanisms as mentioned earlier. The architecture we propose is specifically designed to be robust against congestion and traffic dynamics. To accomplish this objective, we integrate RSSI with performance metrics evaluated in real-time indicating the current state of entire forwarding paths. Informed by the computed metrics, the framework, then, implements a flexible make-before-break handover mechanisms which dynamically selects the best path.

The performance of each path from the GCS to the UAV is measured using beacon messages. Specifically, the GCS periodically generates beacons: small packets containing the generation timestamp and the destination AP information. These beacons are forwarded to all the APs that the GCS can reach through the backbone network.
The UAV monitors all the WiFi channels and capture the broadcast beacons from all the APs in its vicinity.

\begin{figure}[!t]
\centering   \includegraphics[width=.5\textwidth]{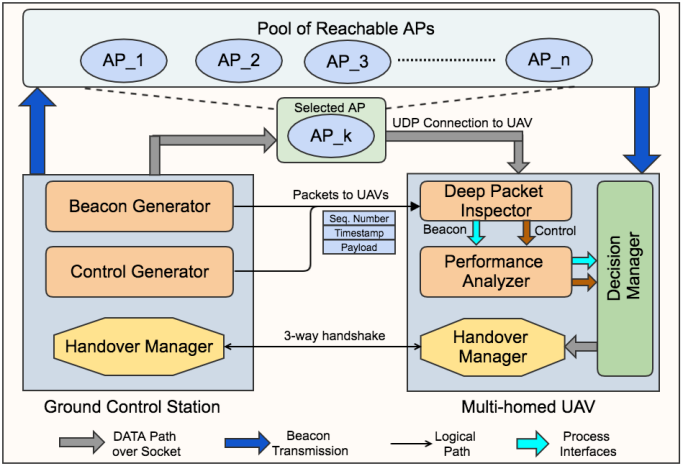}
\vspace{-6mm}
\caption{Proposed flexible and robust communication architecture.}
\vspace{-6mm}
\label{fig:sysarch}
\end{figure}

As illustrated in Fig.~\ref{fig:sysarch}, the framework we propose consist of different functional blocks at the GCS and UAV. The functional blocks at the GCS are: \emph{Control Generator}, \emph{Beacon Generator} and \emph{Handover Manager}. The UAV functional blocks are: \emph{Deep Packet Inspector}, \emph{Performance Analyzer}, \emph{Decision Manager} and \emph{Handover Manager}. In the following, we describe each of these blocks in detail.

\noindent
$\Box$ {\bf{GCS - Control Generator}}: 
This block generates and handles the control messages to be forwarded to the UAV. In the considered case, control messages belong to two classes: heartbeat messages and navigation messages. The former are small messages that are periodically generated so that the UAV can monitor the connection with the GCS. Navigation messages determine the motion of the UAV, and in the considered case contain target GPS coordinates and speed. These messages are defined by the mission control block, which is not explicitly included in the proposed architecture. The Control Generator block adds a generation timestamp and a sequence number to all the control messages. This information is used by the UAV to monitor the quality of the path used to communicate with the GCS in terms of absolute delay and message loss rate.

\vspace{1mm}
\noindent
$\Box$ {\bf{GCS - Beacon Generator}}: 
Note that control messages are routed only through the path currently used to interconnect the GCS to the UAV. Thus, the timestamps and sequence numbers do not provide any information on all the other possible path options. In fact, broadcasting the control messages over the entire backbone network may increase congestion, especially in scenarios with a large number of UAVs. To address this issue, the Beacon Generator periodically generates small messages -- containing a timestamp and a sequence number -- that are broadcasted to all the APs. Note that the UAV does not need to be associated with any specific AP to receive the beacons.

\vspace{1mm}
\noindent
$\Box$ {\bf UAV - Deep Packet Inspector (DPI)}:
This block, implemented at the UAV side, captures all the beacons and control packets. The beacons are collected from all the APs and channels the UAV can receive from, whereas control messages are received only from the currently used path. The block inspects each received packets and creates a data point including the message type, the reception time, the sequence number and the originating AP (MAC address). This information is forwarded to the Performance Analyzer.

\vspace{1mm}
\noindent
$\Box$ {\bf{UAV - Performance Analyzer}}: 
This block receives the data points from the DPI block and determines per-message class packet loss rate and average absolute delay. These performance metrics are measured over a moving time window of duration equal to $\Delta$~seconds. The metrics are forwarded to the Decision Manager, where the moving average measures are used to trigger handover events based on control messages and select the best path based on beacons. In addition to message-related measures, the Performance Analyzer also measures the average RSSI associated with the various interconnected APs.

Note that the duration of the moving window influences the response time and frequency of the framework. On the one hand, a long window better smooths ``noise'', removing small delay and loss peaks, and avoids frequent handover. On the other hand, a short window allows a faster reaction of the framework to congestion. A thorough study on the effect of this parameter on the performance of control delivery and UAV navigation is not included here due to space constraints, and is deferred to future studies.

\vspace{1mm}
\noindent
$\Box$ {\bf{UAV - Decision Manager}}:
The Decision Manager block uses the moving average performance metrics derived by the Performance Analyzer to perform two functionalities: \emph{(a)} Trigger handover to a different AP; and \emph{(b)} Select the best path when a handover event is triggered. In the former functionality, only metrics relative to control messages are used, as handover is necessary only when the QoS of the current path suffers a degradation sufficient to impair the ability of the GCS to control the UAV. The latter functionality considers metrics relative to beacon reception from all the APs, as path selection requires the evaluation of all the feasible paths.

At time instant $t$ the Decision Manager receives moving average beacon delays $D^{i}_{\rm b}(t)$, RSSI $R^{i}_{\rm b}(t)$ and loss rate $L^i_{\rm b}(t)$ corresponding to AP $i$, with $i=1,\ldots,N$, moving average control delay $D_{\rm c}(t)$, RSSI $R_{\rm c}(t)$ and loss rate $L_{\rm c}(t)$.
A handover request is issued at time $t$ if one of the following conditions is satisfied:
\vspace{-1.5mm}
\begin{equation}
 \lambda_1  D_{\rm c}(t) {+} \lambda_2  L_{\rm c}(t) {+} \lambda_3 (R_{\rm max}{-}R_{\rm c}(t))  {>} \Theta;~~
 L_{\rm c}(t){>} \Phi,
\vspace{-1mm}
\end{equation}
where $\lambda_1$, $\lambda_2$ and $\lambda_3$ are positive weights, with $\lambda_1{+}\lambda_2{+}\lambda_3{=}1$, and $R_{\rm max}$ is the maximum RSSI index. $\Theta$ and $\Phi$ are positive thresholds. The first condition corresponds to a general degradation of the current path. In addition to the first condition, we include in the framework an urgent handover mechanism to recover from harsh events in which the connection with the current AP is abruptly severed. Specifically, if the number of heartbeats received in the window is below a certain threshold, the handover manager is immediately notified. This event corresponds to the second condition.

If a handover request is issued, the Decision Manager computes
the metric
\begin{equation}
W_{i}(t) =  \gamma_1  D^{i}_{\rm b}(t) + \gamma_2  L^{i}_{\rm b}(t) + \gamma_3 (R_{\rm max}{-}R^{i}_{\rm b}(t)),
\end{equation}
for all the APs $i=1,2,...,N$, where $\gamma_1$, $\gamma_2$ and $\gamma_3$ are positive weights, with $\gamma_1{+}\gamma_2{+}\gamma_3{=}1$ and $\Theta$ is a positive threshold defining the minimum accepted performance.
The path is selected as
\vspace{-2mm}
\begin{equation}
	k = \arg \min_{i}\{W_{i}(t)\}.
\vspace{-2mm}
\end{equation}
Thus, the decision manager selects the $k^{th}$ AP as the new control path if a handover request is triggered. In this case, the Decision Manager forwards to the UAV Handover Manager the handover request and the index of the new selected AP.

\begin{figure}[!t]
   \centering
   \includegraphics[width=0.5\textwidth]{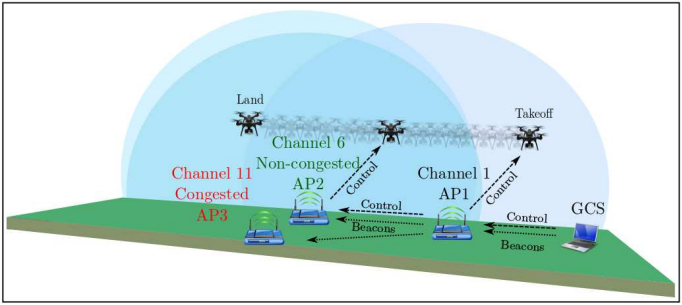}
    \vspace{-6mm}
    \caption{Topology of the experimental setup.}
    \vspace{-6mm}
    \label{fig:linear-sense}
\end{figure}

\noindent
$\Box$ {\bf{Handover Manager}}: 
The handover manager block is located both at the GCS and UAV sides, and implements a 3-way handshake mechanism. The GCS maintains a data structure thats maps the connected UAVs to their corresponding IP addresses. Each UAV keeps track of the GCS's IP address which we assume to be fixed for the duration of the mission. If a handover request is triggered by decision manager, the handover manager at the UAV associates itself with the AP provided by the decision manager. The Dynamic Host Configuration Protocol (DHCP) server at the new AP provides an IP address to the UAV. To ensure make-before-break handover, at this point the UAV doesn't disassociate itself from the old AP and keeps receiving the control messages through that. Now, the UAV initiates handover by sending a handover request message to the GCS via both the APs to maximize the reception probability at GCS. The handover request message contains the UAV's new IP address and the information of new AP. Upon receiving this handover request message, the GCS sends an approve message and note the information received by the request message. Upon reception of approve message, the UAV completes the 3-way handshake by sending ACK message. After a successful handover, the GCS station forwards the control messages over the new path and the UAV disassociates itself from the old AP to save energy.

\section{Experimental Setup and Numerical Results}
\label{numresult}
We assess the performance of the proposed architecture and framework by means of real-world experiments.

\subsection{Experimental Setup}
In the considered setup, the backbone network is composed of three paths through three APs connected to a GCS. The topology is illustrated in Fig.~\ref{fig:linear-sense}: the GCS is connected with AP1, and AP2 and AP3 are at two hop distance from the GCS. All the three APs operate on non-overlapping channels. 

We use Raspberry Pi (RPi) to create the APs using hostapd and all the APs operate according to the IEEE 802.11b standard. The APs communicate with each other via static routing. The GCS, which runs on a laptop, generates a beacon every $200$~ms and a heartbeat message every $500$~ms. The frequency of beacons and heartbeat messages can be increased or decreased based on the observed coherence time of the system.
UDP is used as transport layer for both beacons and control messages. To synchronize the clocks among the UAV and GCS, we use the Network Time Protocol (NTP) with the GCS set as the NTP server.

The UAV is a 3DR solo quad-copter connected to an on-board RPi via a serial link. The RPi is enclosed in a custom 3D printed case. We used the dronekit helper library to communicate with the Pixhawk 2.0 flight controller embedded in the UAV. The RPi is connected with $5$ external wireless dongles: $3$ dongles are used in monitor mode to capture the beacons in WiFi channel $1$, $6$, and $11$, and the remaining $2$ dongles are used to support the make-before-break handover. tcpdump is used to capture the beacons.

The UAV operates in Guided mode, which uses GPS to navigate to way-points (latitude and longitude coordinates). The GCS transmits a predefined series of messages instructing the UAV to navigate to checkpoints. We consider two congestion scenarios using the Iperf utility: \emph{Scenario 1:} a continuous stream of traffic is routed through AP3 path; and \emph{Scenario 2:} the competing traffic stream is alternated between AP2 and AP3 path.

\subsection{Numerical Results}

\begin{figure}[!t]
\centering   \includegraphics[width=.45\textwidth]{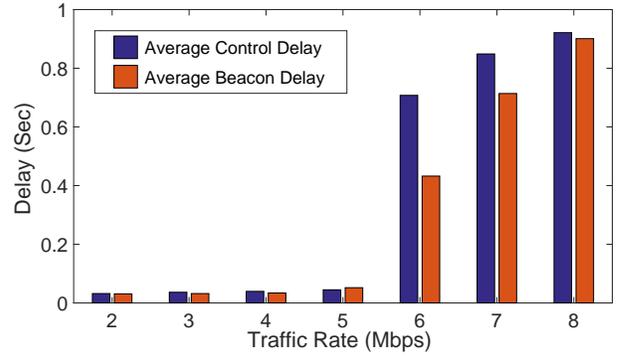}
\vspace{-2mm}
\caption{Impact of congestion on beacon and control messages delay. The overall injected traffic volume is equal to $7$~Mbps.}
\vspace{-2mm}
\label{fig:delay_traffic}
\end{figure}

Fig.~\ref{fig:delay_traffic} shows the average beacon and control delay for different levels of traffic injected at AP3 path, with which the UAV is connected. The maximum achievable traffic volume traversing an individual AP is equal to $8$~Mbps. It is apparent how congestion affects delay as it approaches the maximum supported rate. We observe that in the congestion region, control messages suffer a larger degradation. This is most likely due to the larger size of control packets with respect to beacons. The beacons utilize only a small fraction (approximately 0.045\%) of the total achievable throughput.

\begin{figure}[!t]
\centering   \includegraphics[width=.5\textwidth]{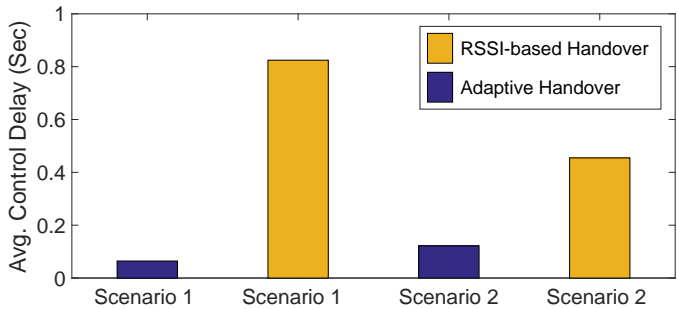}
\vspace{-5mm}
\caption{Average control delay obtained by the handover strategies in Scenario 1 and 2.}
\vspace{-2mm}
\label{fig:handover_delay}
\end{figure}

Fig.~\ref{fig:handover_delay} depicts the average delay of control messages achieved by different handover strategies in Scenario 1 and 2. The overall injected traffic volume is equal to $7$~Mbps. We test an RSSI-based handover strategy against our adaptive handover framework. It can be observed the considerable reduction in delay granted by the proposed framework. Note that the delay in the RSSI-based handover strategy halves in Scenario 2 with respect to Scenario 1. In fact, in the former the congestion is equally spread through the APs, with the UAV connected to one of them in periods uncorrelated with respect to the congestion level. The delay obtained using the proposed technique increases in Scenario 2, where the UAV is forced to shift between AP2 and AP3, suffering a delay penalty due to congestion detection and the establishment of the new forwarding connection for control messages.

\begin{figure}[t]
\centering   \includegraphics[width=.5\textwidth]{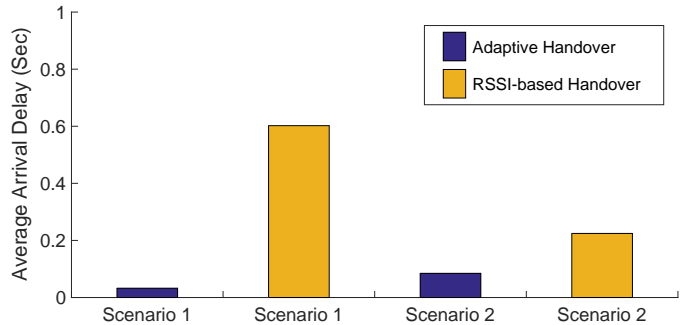}
\vspace{-5mm}
\caption{Average arrival delay to meet the predefined checkpoints in Scenario 1 and 2.}
\vspace{-6mm}
\label{fig:relative_delay}
\end{figure} 
We observed that RSSI and delay are largely uncorrelated in the considered scenario. In fact, although RSSI influences the maximum transmission rate of the direct wireless link between the APs and the UAV, beacon (and control) messages are small messages with small transmission time. Congestion at the AP's buffer or intermediate router results in a delayed forwarding of the packets. Thus handover necessarily needs to use additional information collected by routing packets through the possible paths connecting the GCS to the UAV. Note that RSSI may play a bigger role in determining the overall delay when heavier data streams, \emph{e.g.}, telemetry, are considered.

In addition to the measurement of network performance metric, we illustrate the beneficial impact of the proposed technique on UAV control. In this experiment, we define a sequence of instructions that guide the UAV through a series of waypoints (GPS coordinates).
Fig.~\ref{fig:relative_delay} depicts the average delay in reaching each individual checkpoint granted by the handover techniques with respect to a case with no congestion in Scenario 1 and 2. The reduced time needed to deliver the control messages from the GCS to the UAV granted by the proposed technique results in a reduced delay in reaching the waypoints with respect to RSSI-based handover. Again, we notice the same trend where Scenario 2 mitigates congestion in RSSI-based handover and penalizes the proposed technique due to the more frequent handover events triggered by the alternated traffic injection.

\begin{figure}[t]
   \centering
   \includegraphics[width=.45\textwidth]{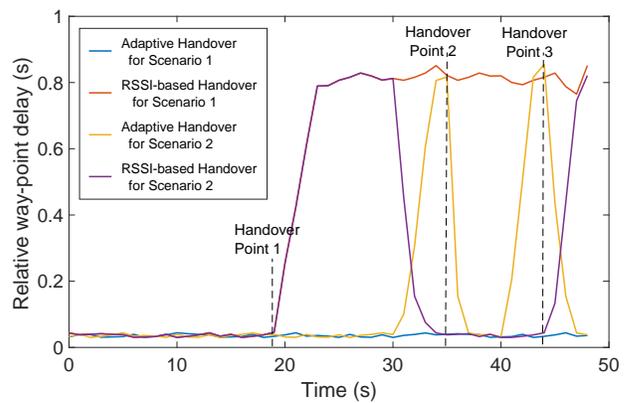}
   \vspace{-4mm}
   \caption{Temporal trace of the relative delay to meet the predefined checkpoints in Scenario 1 and 2.}
   \vspace{-4mm}
   \label{fig:rel_way_delay}
\end{figure}

Fig.~\ref{fig:rel_way_delay} shows the temporal traces of the relative delay. It can be observed that RSSI-based handover incurs periods of large delay when congestion affects the AP used to communicate with the GCS. The proposed technique has short delay peaks corresponding to handover events.

\vspace{-2mm}
\section{Acknowledgments}
\label{sec:ACK}

The architecture proposed and studied in this paper is inspired by the platform and mission description provided at the DARPA Hackfest on Software Defined Radios. The authors of this paper participated as a team to the event. This work was partially supported by the NSF under grant IIS-1724331.
\vspace{-1mm}
\section{Conclusions}
\label {conclusion}
In this paper, we developed a robust communication framework for UAVs operating in congested urban environments. The framework builds a multi-hop multi-path infrastructure used to connect the Ground Control Station to the UAVs. The paths are dynamically selected based on beacon messages that are periodically broadcasted over all the paths.
Experimental results shows that the proposed framework considerably improves the reliability and effectiveness of control against local congestion.

\vspace{-1mm}
\bibliographystyle{IEEEtran}
\bibliography{main}

\end{document}